\documentclass{pas}
\pdfpagewidth=\paperwidth
\pdfpageheight=\paperheight

\usepackage{multirow}
\usepackage{graphicx}
\usepackage{amsmath}
\usepackage{bm}

\begin{document}

\lefttitle{Publications of the Astronomical Society of Australia}
\righttitle{Lu and Blackman}

\jnlPage{1}{4}
\jnlDoiYr{2026}
\doival{10.1017/pasa.xxxx.xx}

\articletitt{Research Paper}

\title{Effects of Rotation on the Gravitational Momentum Transfer in Neutron-Star Kicks and Implications for Spin-Kick Alignment}


\author{\gn{Yiming} \sn{Lu} $^{1}$ and \gn{Eric G.} \sn{Blackman}$^{1}$}
\affil{$^1$Department of Physics and Astronomy, University of Rochester, Rochester, NY 14627, USA 
}

\corresp{Eric G. Blackman, Email: blackman@pas.rochester.edu}

\citeauth{Author1 C and Author2 C, an open-source python tool for simulations of source recovery and completeness in galaxy surveys. {\it Publications of the Astronomical Society of Australia} {\bf 00}, 1--12. https://doi.org/10.1017/pasa.xxxx.xx}

\history{(Received xx xx xxxx; revised xx xx xxxx; accepted xx xx xxxx)}

\begin{abstract}
Neutron stars are often born with large recoil velocities, or natal kicks, whose physical origin remains an open question in core-collapse supernova theory.  Observations suggest that the spin-kick angle distribution is not isotropic but skewed toward a spin-kick vector alignment.  
One possible kick mechanism is the gravitational tug-boat effect, in which anisotropic ejecta gravitationally accelerate the proto-neutron star over a timescale of seconds after shock revival, although the long-term importance of this effect relative to other hydrodynamic forces remains debated.
Previous derivations of the tug-boat mechanism do not include the effect of initial stellar rotation. Here we derive a  minimalist extension  to assess how rotation of  the expanding  asymmetric mass distribution influences  spin-kick alignment of gravitational momentum transfer.
We show that the spin-kick angle is determined by the product of two factors, one that depends on the ratio of shock expansion time to the rotation period and the other which depends on the orientation of the asymmetric mass distribution with respect to the spin-axis. 
For fast enough rotation, the first factor amounts to axially averaging out non-axisymmetry thereby suppressing the perpendicular tug and leaving only a spin-aligned force. However, the rotation speed required for this effect would be unrealistically large unless magnetic fields could transport angular momentum from the core to the outflow efficiently. Spin-kick alignment by the tug-boat  mechanism would otherwise require the second factor, namely a preferentially spin-aligned mass flux asymmetry.  The rotational-averaging framework developed here suggests that, in general, for any kick mechanism that is not sourced by rotation, including rotation will have some tendency to spin-align the kick, but reduce the kick magnitude.

\end{abstract}


\maketitle

\section{Introduction}

Neutron stars are observed to move with large space velocities after birth,
typically  hundreds of kilometers per second, with some  reaching  $1000~{\rm km~s^{-1}}$ \citep[e.g.,][]{1996AIPC..366...25B,1996PhRvL..76..352B,1998Natur.393..139S,2002ApJ...568..289A,2005MNRAS.360..974H}.
Momentum conservation dictates that these natal kicks are a direct signature of  asymmetry in core-collapse supernova explosions.  
In addition to the kick magnitude, the direction of the kick relative to the neutron-star spin axis provides an important clue to the underlying explosion dynamics.
A tendency toward spin-kick alignment has been identified, although the observational sample remains limited and projection effects are important \citep[e.g.,][]{2005MNRAS.364.1397J, 2007ApJ...660.1357N, 2012MNRAS.423.2736N, 2025PASA...42..106B}.

Several mechanisms have been proposed to explain neutron-star kicks.
Anisotropic neutrino emission can in principle produce a recoil because neutrinos carry away most of the gravitational binding energy of the proto-neutron star.
However, producing a large enough neutrino anisotropy generally requires extremely strong magnetic fields and special conditions \citep[e.g.,][]{1998ApJ...505..844L}.
A more hydrodynamical possibility is that asymmetric ejecta carry linear momentum in one direction, causing the neutron star to recoil in the opposite direction by momentum conservation \citep{2010PhRvD..82j3016N}.
Early simulations suggested that this prompt hydrodynamical impulse alone may be too weak to account for the largest observed kicks \citep{1994A&A...290..496J}.

One proposed 
mechanism is the gravitational tug-boat effect \citep{2006A&A...457..963S,2012MNRAS.423.1805N,2013A&A...552A.126W,2017ApJ...837...84J}
in which asymmetric ejecta gravitationally pull on the proto-neutron star after shock revival.
Because this force can persist for seconds, the kick can, in principle, grow long after the initial explosion asymmetry is established.
The largest kicks are then associated with strong large-scale asymmetries, especially dipolar ejecta distributions. We focus on this gravitational momentum transfer in this paper, although whether other hydrodynamic forces may become dominant on timescales beyond those currently accessible to simulations remains debated \citep{2022MNRAS.517.3938C,2024ApJ...963...63B}. Other proposed mechanisms include kicks produced by stochastic, asymmetric jet activity \citep{2009MNRAS.396.1783P,2024OJAp....7E..12S,2025RAA....25d5008B,2025ApJ...992..190S}. 

Previous  detailed studies of the gravitational momentum transfer in the tug-boat mechanism did not include progenitor rotation.
When the progenitor rotates, core-collapse explosions rotate in  a preferred sense, with  typical rotation rates exceeding that induced by stochastic torques \citep{2013A&A...552A.126W}. Any asymmetric mass distribution would also rotate as it expands. This raises two natural questions: 1. How does the rotation of an asymmetric ejecta pattern modify the accumulated gravitational 
kick? 2. Can this provide a  
physical explanation for the observational skew toward spin-kick alignment of neutron stars?

In this paper, we develop a minimalist analytic model to address these questions.
In Section 2, we discuss the basic non-rotating tug-boat mechanism and further support the validity of the simplified model by considering back-reaction. We add rotation in section 3, where we show that the angle between spin and kick depends on two factors, one associated with the rotation speed and the other on the initial mass asymmetry.
In section 4 we discuss the implications for spin-kick alignment and the possible role of magnetic fields.  We present the conclusions, limitations, and targets for further work in section 5.

\section{The Gravitational Tug-Boat Mechanism Revisited}
\subsection{The basic model}\label{2.1}
\begin{figure}
    \centering
    \includegraphics[width=\columnwidth,trim=3.5in 2.0in 4.0in 1.0in,clip]{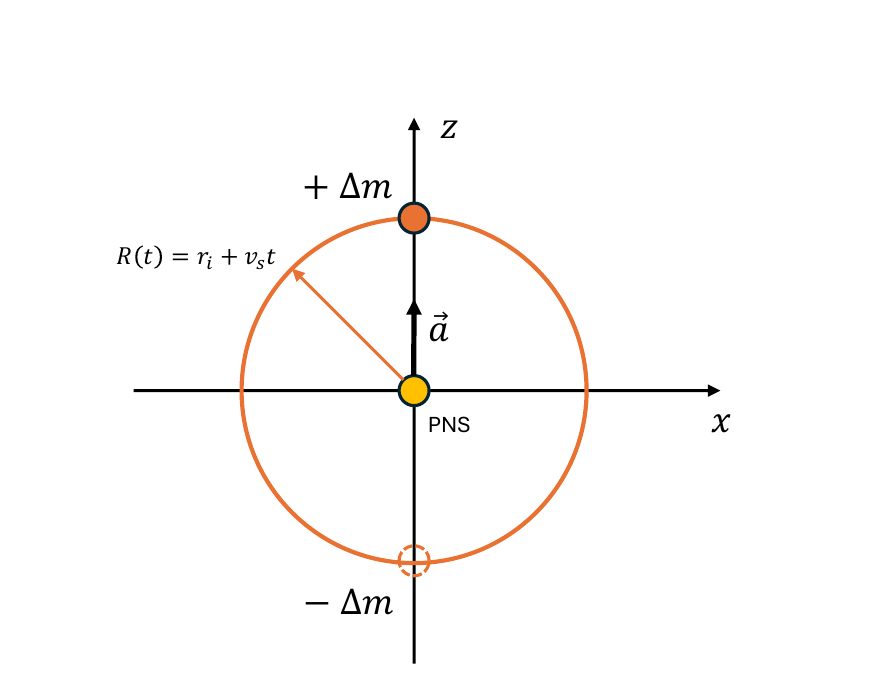}
    \caption{
    Schematic illustration of the non-rotating gravitational tug-boat model. 
    }
    \label{fig:tugboat}
\end{figure}

The basic non-rotating gravitational tug-boat picture can be illustrated with a simple toy model following \citet{2013A&A...552A.126W}.  The model assumes an initial asymmetry in the ejecta mass distribution, which can be represented by an effective mass excess $\Delta m$ at radius $R$ from the proto-neutron star, together with a corresponding mass deficit $-\Delta m$ on the opposite side, as shown in Figure~\ref{fig:tugboat}.
Here, $\Delta m$ is an effective dipolar mass asymmetry that represents the equivalent gravitational force of the asymmetric ejecta on the proto-neutron star, and therefore depends on the ejecta distribution and shape.
The gravitational acceleration of the neutron star is then approximately
\begin{equation}
    a_{\rm NS}
    \simeq
    \frac{2G\Delta m}{R^2}.
    \label{eq:basic_acceleration}
\end{equation}
If the ejecta expand outward with a constant speed $v_s$, then
\begin{equation}
    R(t) = r_i + v_s t,
    \label{eq:constant_expansion}
\end{equation}
where $r_i$ is the initial shock radius at the onset of the tug-boat phase.
For the non-rotating case, we can assume that the asymmetry direction is fixed and aligned with the final kick direction. Then the accumulated kick velocity magnitude is:
\begin{equation}
    v_{\rm kick}
    =
    \int_0^\infty
    \frac{2G\Delta m}{(r_i+v_s t)^2}\,dt
    =
    \frac{2G\Delta m}{r_i v_s}.
    \label{eq:tugboat_velocity}
\end{equation}
This estimate captures the main reason why the tug-boat mechanism can be efficient: the acceleration is modest instantaneously but can accumulate during the ejection. 

If we choose a polar axis a priori,
then the kick velocity vector will be at some angle to the polar axis, depending on the asymmetric mass distribution. For arbitrary mass asymmetry, we can always approximate the resultant force by a pair of effective perturbed point masses equating their net gravitational force to that  associated with the  mass asymmetry. This justifies the utility of the simplified model.

The above calculation shows the effect of mass asymmetry. However, the tug-boat mechanism is also affected by velocity asymmetry within the ejecta: slow-moving clumps can gravitationally pull on the neutron star for a longer time than faster-moving material. In fact, the configuration shown in Figure~\ref{fig:tugboat} requires a non-zero initial net momentum asymmetry. This is possible due to the neutrino emission and accretion before the gravitational force becomes dominant \citep[e.g.,][]{2006A&A...457..963S,2024Ap&SS.369...80J}. But the effectiveness of the gravitational tug-boat mechanism actually does not rely on any pre-existing momentum asymmetry. Assume that initially both the proto-neutron star and the ejecta have zero net momentum, but one side of the ejecta is more massive and expands more slowly, causing it to exert a stronger and longer-lasting gravitational pull on the proto-neutron star than the material on the opposite side. As a result, momentum is transferred gravitationally to the proto-neutron star. In this scenario, the kick velocity of the proto-neutron star and the momentum asymmetry of the ejecta grow together.

We explicitly show below that velocity and mass asymmetries can contribute at the same order when the initial momentum asymmetry is zero (or small). Let the clump with mass $m+\Delta m$ move with velocity $v-\Delta v$, while the opposite clump with mass $m-\Delta m$ moves with velocity $v+\Delta v$. 
The final kick velocity is given by the difference between the two gravitational contributions, each computed separately from  equation (\ref{eq:tugboat_velocity}), giving
\begin{equation}\label{eq:kick_velasymmetry}
\begin{aligned}
    v_{\rm kick}
    &=
    \frac{G(m+\Delta m)}
         {r_i(v-\Delta v)}
    -
    \frac{G(m-\Delta m)}
         {r_i(v+\Delta v)}
    \\
    &\approx
    \frac{2G}{r_i v^2}
    \left(
        m\Delta v
        +
        v\Delta m
    \right).
\end{aligned}
\end{equation}
Because the initial net momentum is assumed to be zero,
\begin{equation}
    (m+\Delta m)(v-\Delta v)
    =
    (m-\Delta m)(v+\Delta v).
\end{equation}
To first order, we have
\begin{equation}
    m\Delta v
    =
    v\Delta m, 
\end{equation}
showing that the contribution from the velocity asymmetry is  exactly the same as that from the mass asymmetry in equation~(\ref{eq:kick_velasymmetry}). Therefore, the final kick velocity should be
\begin{equation}\label{eq:kick_vel_corrected}
    v_{\rm kick}
    =
    \frac{4G\Delta m}{r_i v_s}.
\end{equation}
Equation~(\ref{eq:kick_vel_corrected}) shows that a correction factor of 2 should be introduced into Equation~(\ref{eq:tugboat_velocity}) when the initial net momentum is zero. 
In practice, we can mathematically consider only the initial mass difference 
and multiply the final result by 2 to obtain the net physical contributions from both mass and velocity differences. This calculation also shows that, even if the initial momentum asymmetry is insufficient to explain the final neutron-star kick, additional momentum can be transferred gravitationally from the initial mass and velocity asymmetries.

The typical values of $\Delta m$ and $v_s$ are $10^{-2}\,M_\odot$ and $10^{4}~{\rm km~s^{-1}}$, respectively \citep{2017ApJ...837...84J}. The ejecta velocity is of the same order as the shock velocity:
\begin{equation}
\begin{aligned}
    \bar{v}_{\rm ej}
    &=
    \sqrt{
        \frac{2f_{\rm kin}E_{\rm tot}}
             {M_{\rm ej}}
    }
    \\
    &\approx
    10^{4}\,{\rm km\,s^{-1}}
    \left(
        \frac{f_{\rm kin}}{0.1}
    \right)^{1/2}
    \left(
        \frac{E_{\rm tot}}
             {10^{51}\,{\rm erg}}
    \right)^{1/2}
    \\
    &\qquad\times
    \left(
        \frac{M_{\rm ej}}
             {0.1\,M_\odot}
    \right)^{-1/2},
\end{aligned}
\end{equation}
where $f_{\rm kin}\approx0.1$ \citep{2017ApJ...837...84J} is the fraction of the explosion energy carried by the kinetic energy of the ejecta. Therefore, it is reasonable to relate the expansion velocity of the perturbed mass to the shock expansion velocity.

The value of $r_i$ is more subtle.  It should be interpreted as the shock radius at which the tug-boat mechanism becomes effective. Therefore, taking $r_i$ to be the shock radius at the onset of shock revival can overestimate the kick velocity. From  figure 1 in \cite{2012MNRAS.423.1805N}, we  see that the kick velocity grows significantly only after about $0.2\,{\rm s}$. As the shock moves outward, it sweeps up  material, and the ejecta mass increases. The gravitational acceleration becomes important only when the asymmetric mass is sufficiently large.

Momentum conservation is not violated in the simple model because the ejecta can have a very large velocity at the early stage. Once it mixes with the exterior material, its velocity decreases and its mass increases while its momentum remains approximately unchanged. Once a sufficient amount of mass has accumulated, the tug-boat model starts to work. We therefore adopt $r_i\approx10^{4}\,{\rm km\,s^{-1}} \times 0.2\,{\rm s}=2000\,{\rm km}$.

Taking these representative values, equation~(\ref{eq:kick_vel_corrected}) gives:
\begin{equation}
\begin{aligned}
    v_{\rm kick}
    &=
    \frac{4G\Delta m}{r_i v_s} \\
    &\simeq
    270\,{\rm km\,s^{-1}}
    \left(
        \frac{\Delta m}{10^{-2}\,M_\odot}
    \right)
    \left(
        \frac{r_i}{2000\,{\rm km}}
    \right)^{-1} \\
    &\quad\times
    \left(
        \frac{v_s}{10^4\,{\rm km\,s^{-1}}}
    \right)^{-1},
\end{aligned}
\end{equation}
which is a reasonable magnitude for the saturated kick velocity.

\subsection{Back-reaction: a mass-changing model}
\label{2.2}
In the simplified derivation of the tug-boat phase above, the proto-neutron star and the effective mass asymmetry were treated as fixed, while the back-reaction of the proto-neutron star's gravitational force on the ejecta was neglected. To estimate the correction associated with this back-reaction, we allow the effective mass asymmetry $\Delta m$ to vary with time and impose momentum conservation of the system.

This provides a simple way to represent the evolving fluid distribution of the ejecta. Local density perturbations can be modified by pressure gradients and mixing with the surrounding flow. More importantly, as defined above, $\Delta m$ is an effective quantity determined by the global density distribution and shape of the asymmetric ejecta, not just the total relative hemispheric mass. It can  evolve as the ejecta expand and deform, for example, if one hemisphere of the mass distribution evolves from spherically symmetric to a teardrop shape.
Allowing $\Delta m=\Delta m(t)$ thus provides a simple phenomenological description of the evolving gravitational asymmetry while accounting for the back-reaction required by momentum conservation.

We therefore adopt a mass-changing toy model to estimate the effect of back-reaction within the simplified tug-boat framework.
The equation of motion for the neutron star is
\begin{equation}
    M\dot{v}_{\rm NS}
    =
    \frac{2G\Delta m(t)M}{R^2},
\end{equation}
where we ignored any pressure gradient force on the neutron star  compared to the gravitational force. Assuming that neutrino-induced momentum transfer is subdominant during this phase, approximate momentum conservation gives
\begin{equation}
\begin{aligned}
    Mv_{\rm NS}
    +
    2\Delta m(t)v_s
    &=
    P_0
    \\
    &=
    2\Delta m(0)v_s.
\end{aligned}
\label{11}
\end{equation}
Here we do not explicitly include the velocity asymmetry as discussed in Section~\ref{2.1}; its contribution can be approximately accounted for later by the factor of two derived there.

Assuming that $v_s$ is constant and that the displacement of the neutron star is negligible,
\begin{equation}
    \dot{R}
    =
    v_s.
\end{equation}

Taking the time derivative of the momentum equation gives
\begin{equation}
    \frac{{\rm d}\Delta m}{{\rm d}t}
    =
    -\frac{GM}{R^2v_s}\Delta m.
\end{equation}

The solution for the mass asymmetry is
\begin{equation}
\begin{aligned}
    \Delta m(t)
    &=
    \Delta m(0)
    \exp
    \left[
        \frac{GM}{v_s^2}
        \left(
            \frac{1}{R}
            -
            \frac{1}{r_i}
        \right)
    \right].
\end{aligned}
\label{14}
\end{equation}

Using equation (\ref{14}) in equation (\ref{11}), the NS velocity is then
\begin{equation}\label{eq:v_ns_of_t}
\begin{aligned}
    v_{\rm NS}(t)
    &=
    \frac{2\Delta m(0)v_s}{M}
    \left\{
        1
        -
        \exp
        \left[
            \frac{GM}{v_s^2}
            \left(
                \frac{1}{R}
                -
                \frac{1}{r_i}
            \right)
        \right]
    \right\}.
\end{aligned}
\end{equation}

As $t\rightarrow\infty$ and $R\rightarrow\infty$,
\begin{equation}
\begin{aligned}
    \Delta m(t\rightarrow\infty)
    &=
    \Delta m(0)
    \exp
    \left(
        -\frac{GM}{v_s^2r_i}
    \right)
    \\
    &=
    \Delta m(0)
    \exp
    \left(
        -\frac{v_\star^2}{v_s^2}
    \right),
\end{aligned}
\end{equation}
and
\begin{equation}
\begin{aligned}
    v_{\rm NS}(t\rightarrow\infty)
    &=
    \frac{2\Delta m(0)v_s}{M}
    \left[
        1
        -
        \exp
        \left(
            -\frac{GM}{v_s^2r_i}
        \right)
    \right]
    \\
    &=
    \frac{2\Delta m(0)v_s}{M}
    \left[
        1
        -
        \exp
        \left(
            -\frac{v_\star^2}{v_s^2}
        \right)
    \right],
\end{aligned}
\label{17}
\end{equation}
where we define
\begin{equation}
    v_\star
    =
    \sqrt{\frac{GM}{r_i}}.
\end{equation}
$v_\star$ is of the same order as the escape velocity at $r_i$, although it is not identical to the escape velocity. When $v_s\gg v_\star$, this expression reduces to the previous result obtained without including the back-reaction:
\begin{equation}
\begin{aligned}
    v_{\rm NS}(t\rightarrow\infty)
    &\approx
    \frac{2\Delta m(0)v_s}{M}
    \frac{v_\star^2}{v_s^2}
    \\
    &=
    \frac{2\Delta m(0)v_s}{M}
    \frac{GM}{v_s^2r_i}
    \\
    &=
    \frac{2G\Delta m(0)}{v_sr_i},
    \label{19}
\end{aligned}
\end{equation}
while
\begin{equation}
    \Delta m(t\rightarrow\infty)
    \approx
    \Delta m(0).
\end{equation}

\begin{figure}
    \centering
    \includegraphics[width=\columnwidth]{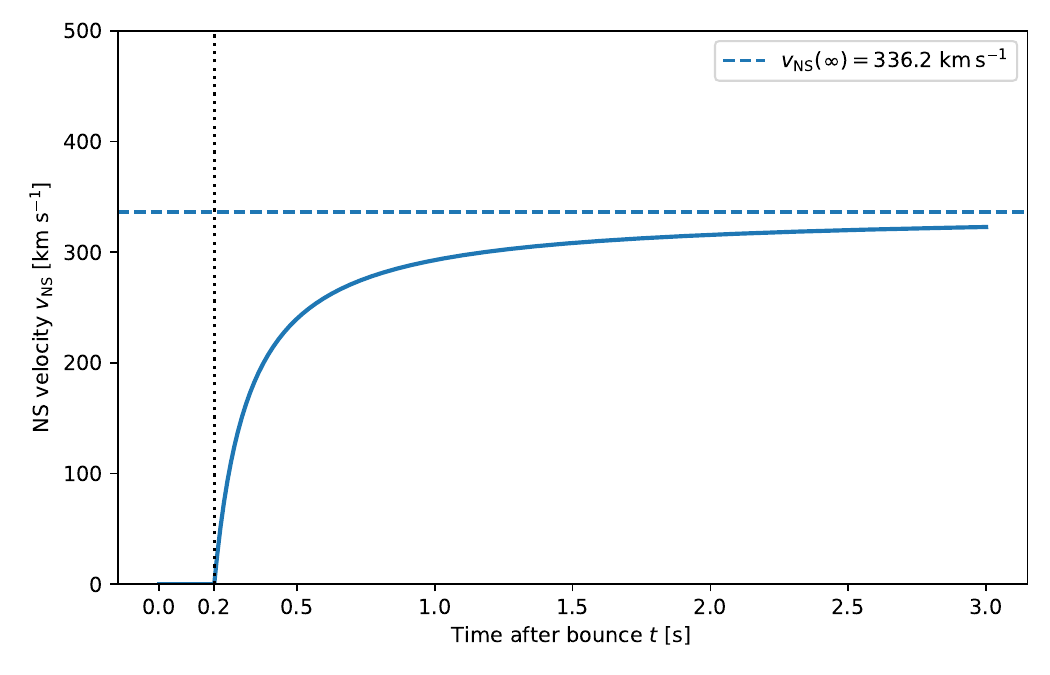}
    \caption{Time evolution of the NS velocity predicted by Equation~(\ref{eq:v_ns_of_t}), including the additional correction factor of 2 as discussed below equation (\ref{eq:kick_vel_corrected}). The fiducial parameters are $r_i = 2000\,\mathrm{km}$, $v_s = 10^4\,\mathrm{km\,s^{-1}}$, $M = 1.5\,M_\odot$, and $\Delta m(0) = 2\times10^{-2}\,M_\odot$. The time origin is shifted to $t=0.2\,\mathrm{s}$, corresponding to the onset of the tug-boat phase.
    }
    \label{fig:v_ns}
\end{figure}

Using the representative values discussed in Section~\ref{2.1} and $M=1.5M_{\odot}$, we have
\begin{equation}
    \left(
        \frac{v_\star}{v_s}
    \right)^2
    \approx
    0.995.
    \label{21}
\end{equation}
So the ratio of  equations (\ref
{17}) to (\ref{19}) is then
\begin{equation}
\begin{aligned}
    \frac{
        v_{\rm kick}
        \text{ (mass-changing model)}
    }{
        v_{\rm kick}
        \text{ (no back-reaction)}
    }
    &=
    \frac{
        1-\exp(-v_\star^2/v_s^2)
    }{
        v_\star^2/v_s^2
    }
    \\
    &\approx
    0.63.
\end{aligned}
\label{22}
\end{equation}
Therefore, including the back-reaction changes the final kick velocity only by a factor of order unity, so the  model without back-reaction in Section~\ref{2.1} remains valid at the order-of-magnitude level. This quantitative correction is sensitive to $r_i$ for values much smaller than our adopted value of  $r_i=2000$ km, although any larger $r_i$   makes the ratio in equation (\ref{22}) even closer to unity. The time evolution of the NS velocity is shown in Figure~\ref{fig:v_ns}.

The decrease of $\Delta m(t)$ in this simple model can be interpreted as a partial weakening of the mass asymmetry as the ejecta expand. Importantly, the asymmetry is not completely erased, but approaches a finite residual value at late times. This is qualitatively consistent with what is expected for a fluid dynamical evolution, but quantifying the exact dynamics requires solving the full 3-D hydrodynamic equations. While we cannot be sure that the back-reaction effect would remain as modest when fully realistic dynamics are included, our simplified model in Section~\ref{2.1} suggests that the effect is not large at the order-of-magnitude level.

\section{Minimalist Rotating Tug-Boat Model} \label{sec:rotate_model}
We now consider a simple rotating version of the tug-boat mechanism. 
For the rotating case, the effective mass perturbation may spiral outward around a prescribed rotation axis, causing the transverse gravitational force to partially cancel over time.

\begin{figure*}
    \centering
    \includegraphics[
        width=\textwidth,
        trim=0.5in 2.5in 0.5in 0.3in,
        clip
    ]{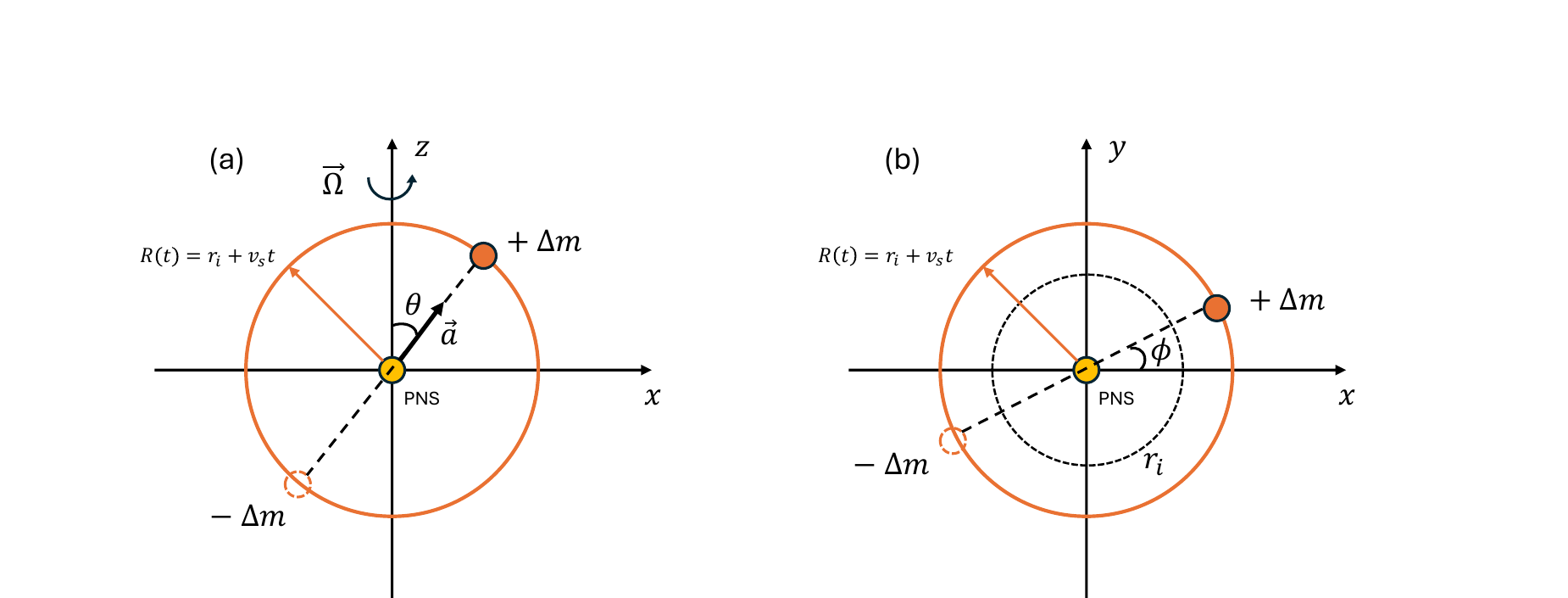}
    \caption{Schematic illustration of the rotating tug-boat model. (a) Side view in the x-z plane. The effective asymmetric point mass is located in polar angle $\theta$, with the rotation axis aligned with the z-axis. Its radial position increases at a constant velocity $v_s$. (b) Top-down view of the model.$\phi$ is the azimuthal angle in the x-y plane and is given by Equation~(\ref{eq:phi}). The perturbed mass spirals outward during the explosion.}
    \label{fig:rotating_model}
\end{figure*}

The model is shown in Figure~\ref{fig:rotating_model}. Let the pre-existing rotation axis define the $z$-axis.
We assume that the effective asymmetric point mass lies at a fixed polar angle $\theta$ from the spin axis, while its azimuthal angle rotates with angular velocity $\Omega(t)$ and expands with constant radial velocity $v_s$ along with the ejecta shell. Assuming that the angular momentum of the effective perturbation is conserved during the expansion, we write
\begin{equation}\label{eq:AM conservatiom}
    j=\Omega(t) R(t)^2=const.
\end{equation}
The rotation rate can then be written as
\begin{equation}\label{eq:omega_time}
    \Omega(t)=\Omega_{i}\left(\frac{r_i}{r_i+v_s t}\right)^2=\Omega_{i}\left(\frac{r_i}{R(t)}\right)^2,
\end{equation}
where $\Omega_i$ is the initial angular velocity and $r_i$ is the initial shock radius at the onset of the tug-boat phase.

The azimuthal angle of the perturbed mass is:
\begin{equation}\label{eq:phi}
\begin{aligned}
    \phi(t)
    &= \int_0^t \Omega(t')\,dt'  \\
    &= \frac{\Omega_i r_i}{v_s}
    \left(1-\frac{r_i}{r_i+v_s t}\right)
     = \frac{\Omega_i r_i}{v_s}
    \left(1-\frac{r_i}{R(t)}\right).
\end{aligned}
\end{equation}
We can define a rotation parameter $\beta$ as
\begin{equation}\label{eq:beta}
    \beta \equiv\frac{\Omega_i r_i}{v_s}=\phi(t\to\infty),
\end{equation}
which is the ratio of the initial rotational velocity and expansion velocity. The parameter $\beta$ also represents the asymptotic azimuthal angle swept out by the perturbed mass.

Given the angle definitions, the acceleration of the neutron star can be written with respect to Cartesian axes as:
\begin{equation}\label{eq:acceleration}
    \vec{a}_{\rm NS} =\frac{2G\Delta m}{R(t)^2}(\sin{\theta}\cos{\phi(t)}\hat{\bm{x}}+\sin{\theta}\sin{\phi(t)}\hat{\bm{y}}+\cos{\theta}\hat{\bm{z}}).
\end{equation}
Equation~(\ref{eq:acceleration}) can be integrated to give the kick velocity. By changing variables to $u(t)=r_i/R(t)$, the integral involved in the $\hat{\bm{x}}$ component of equation~(\ref{eq:acceleration}) is:
\begin{equation}
\begin{aligned}
I_c
&= \int_{0}^{\infty}
\frac{\cos\phi(t)}{R(t)^2}\,dt \\
&= \int_{0}^{1}
\frac{1}{r_i v_s}
\cos\!\left[\beta(1-u)\right]\,du \\
&= \frac{1}{r_i v_s}\frac{\sin \beta}{\beta}.
\end{aligned}
\end{equation}
Other integrals can be similarly calculated. The final result of the kick velocity is:
\begin{equation}\label{eq:kick_v}
    \vec{v}_{\rm kick} = \frac{2G\Delta m}{r_i v_s}(\sin \theta \frac{\sin \beta}{\beta}\hat{\bm{x}}
    +\sin \theta \frac{1-\cos \beta}{\beta}\hat{\bm{y}}+\cos \theta \hat{\bm{z}}).
\end{equation}
The perpendicular and parallel components of the kick velocity are:
\begin{equation}\label{eq:vel}
\begin{aligned}
    v_{\perp}
    &= \frac{2G\Delta m}{r_i v_s}
    \sin\theta\,
    \left|
    \frac{\sin(\beta/2)}{\beta/2}
    \right|, \\
    v_{\parallel}
    &= \frac{2G\Delta m}{r_i v_s}
    \cos\theta .
\end{aligned}
\end{equation}

Using Equation~(\ref{eq:vel}), the angle between the kick direction and the spin axis is
\begin{equation}\label{eq:main}
    \tan\alpha_{\rm sk}
    =
    \frac{v_\perp}{v_\parallel}
    =
    \tan\theta\,
    \left|
    \mathrm{sinc}\left(\frac{\beta}{2}\right)
    \right|.
\end{equation}
Equation (\ref{eq:vel}) shows that for a fixed $\Delta m$, 
the  resulting kick velocity for the rotating case will be  reduced
compared to the non-rotating cases and with a spin-kick angle given by Equation~(\ref{eq:main}).
Numerical simulations of supernovae also find a negative correlation between the final kick velocity and the rotation rate \citep[e.g.,][]{2006A&A...457..963S,2024ApJ...964L..16B}. That may reflect a more general principle of how rotation affects any momentum transfer mechanism, rather than specifically rotation on the tug-boat model.

In this minimalist model, we neglect the gravitational back-reaction of the proto-neutron star on the ejecta.
As in the non-rotating case, we expect this approximation to introduce only a modest quantitative error. A fully self-consistent treatment is more complicated in the rotating case because an initially momentum-symmetric configuration would generally require a reflectionally asymmetric rotation profile across the equatorial plane. 
Assessing the magnitude of the back-reaction therefore requires future numerical calculations based on the fluid equations. Nevertheless, the qualitative conclusion should remain robust: sufficiently rapid rotation averages out a substantial fraction of the transverse acceleration and can consequently enhance spin-kick alignment.

The idea that rotation can average out the transverse component of momentum transfer can in principle be generalized to other forms of momentum transfer, including neutrino emission and other hydrodynamic forces. In our simplified model, the gravitational momentum-transfer rate decays as $t^{-2}$, whereas other processes may have different time dependences.

\section{Implications for spin-kick Alignment}\label{sec:implication}

Equation~(\ref{eq:main}) and its implications are the main results of this paper.  
The theoretical probability distribution of $\theta$, the effective polar angle  of the asymmetric mass distribution,  is likely to be broad because of the complex hydrodynamic evolution, although jet formation may narrow this breadth. Equation~(\ref{eq:main}) shows that the observed distribution of $\alpha_{\rm sk}$ is modulated by a sinc rotational-averaging factor. When this factor is small,  strong spin-kick alignment can occur even if the underlying distribution of $\theta$ is broad.  

\begin{figure}
    \centering
    \includegraphics[width=\columnwidth]{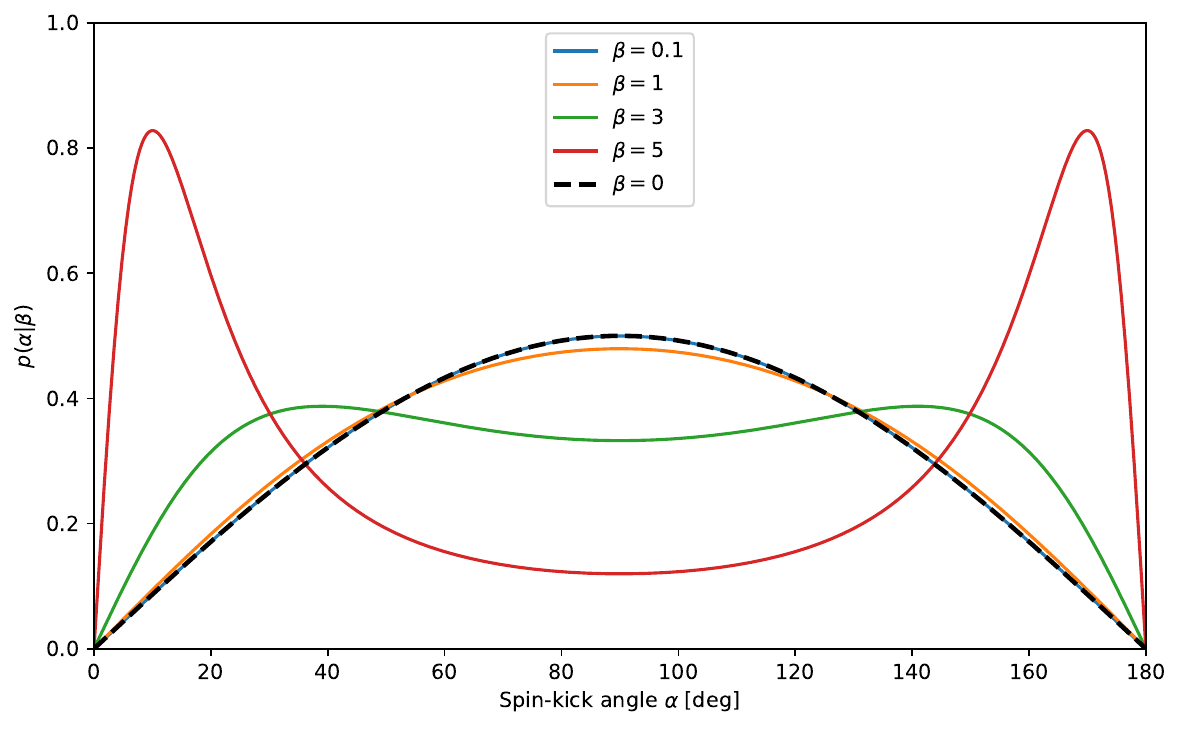}
    \caption{Probability distribution of the spin-kick angle $\alpha$ for different values of $\beta$, calculated from Equation~(\ref{eq:main}). 
    The initial asymmetry direction is assumed to be isotropically distributed, so that $p(\theta)\propto \sin\theta$.
    }
    \label{fig:pdf}
\end{figure}

As an  example, Figure~\ref{fig:pdf} shows how varying $\beta$ reshapes the probability distribution of the spin-kick angle $\alpha$ for a completely random initial asymmetry direction, corresponding to $p(\theta)=\frac{1}{2}\sin\theta$ for $0\leq\theta\leq\pi$.
In the slow-rotation limit,
\begin{equation}
    \beta \ll 1,
\end{equation}
and we have
\begin{equation}
    \mathrm{sinc}\left(\frac{\beta}{2}\right)
    \rightarrow 1,\,
    \alpha_{\rm sk}
    \approx
    \theta.
\end{equation}
The kick direction simply follows the direction of the asymmetric mass, as in the usual non-rotating tug-boat model.
In the rapidly rotating limit,
\begin{equation}
    \beta \gg 1,\,\tan\alpha_{\rm sk} \rightarrow 0,
\end{equation}
meaning that the perpendicular component is reduced by rotational averaging.
For sufficiently large $\beta$, the transverse kick can become much smaller than the parallel kick,
corresponding to extreme alignment between the kick and the spin axis.

\begin{figure}
    \centering
    \includegraphics[width=\columnwidth]{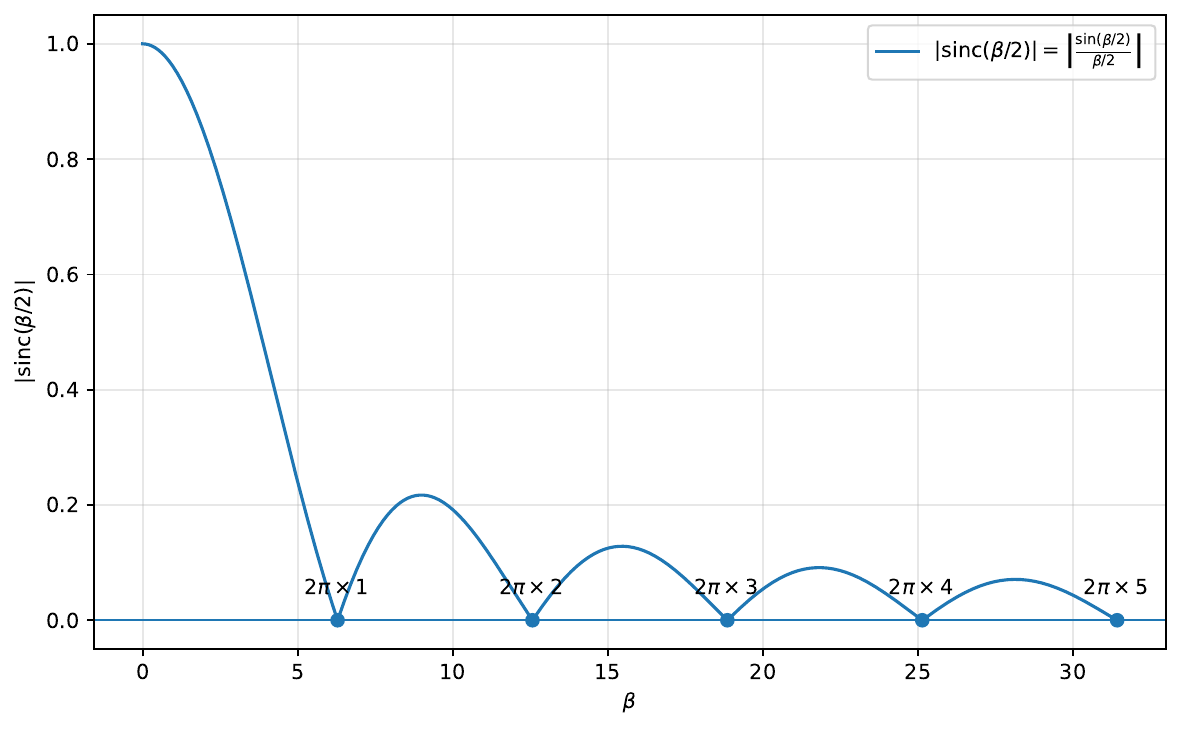}
    \caption{Rotational-averaging factor $|\mathrm{sinc}(\beta/2)|=|\sin(\beta/2)/(\beta/2)|$ as a function of $\beta$. The first zero occurs at $\beta=2\pi$.
    }
    \label{fig:sinc}
\end{figure}

To have a strong spin-kick alignment, $\beta$ does not need to be that large. The first zero point of $\mathrm{sinc}(\beta/2)$ is $\beta_1=2\pi \approx 6$ (Figure~\ref{fig:sinc}), so according to equation~(\ref{eq:main}), even such a moderate $\beta$ can cause subtantial spin-kick alignment.
 However, we must estimate  $\beta$ for real systems. The typical specific angular momentum of the iron core is inferred as $j \lesssim 10^{15}\,\mathrm{cm^2\,s^{-1}}$ \citep{2019MNRAS.486.2238A}. Neglecting angular momentum loss and magnetic redistribution, the rotation rate of the ejecta at the onset of gravitational tug-boat is about $\Omega_i\simeq j/r_i^2\simeq(10^{15}\,\mathrm{cm^2\,s^{-1}})/(2000\,\mathrm{km})^2 \simeq 0.025\,\mathrm{rad\,s^{-1}} $.
Then the rotation parameter associated with typical bulk rotation is
\begin{equation}\label{eq:beta_estimate}
    \beta_{\rm bulk}\simeq
    \frac{0.025\,\mathrm{rad\,s^{-1}}\times 2000\,\mathrm{km}}
    {10000\,\mathrm{km\,s^{-1}}}
    =0.005 .
\end{equation}
This implies that such ordinary progenitor bulk rotation lies in the slow-rotation regime of the present model and is  unable to produce strong spin-kick alignment by itself. Even for extremely rapidly rotating progenitors, $\beta$ may remain modest because the ejecta expansion speed can also be substantially larger. For example, in the 3-D GRMHD simulations of \cite{2026MNRAS.548ag646S}, models with $\Omega_0 \gtrsim 2.0\,\mathrm{rad\,s^{-1}}$ have ejecta expansion speeds of $v_s \gtrsim 15000\,\mathrm{km\,s^{-1}}$. Using their initial rotation profile and assuming angular momentum conservation, we estimate that these models correspond to $\Omega_i \sim 0.1\,\mathrm{rad\,s^{-1}}$.
 The corresponding value of $\beta$ therefore remains below unity even for such rapidly rotating cases.

The parameter $\Omega_i$ in Equation~(\ref{eq:beta}) should be interpreted as the
actual bulk rotation during the tug-boat phase,
not as a pattern speed of a hydrodynamic oscillation mode. This distinction
matters for spiral SASI modes, which can produce rotating non-axisymmetric
patterns in the post-shock flow \citep[e.g.,][]{2003ApJ...584..971B,2007ApJ...656..366B,2017MNRAS.471..914K}
with typical inferred frequencies of order $f_{\rm SASI}\sim100\,\mathrm{Hz}$
\citep{2019MNRAS.486.2238A,2021MNRAS.503.3552A}. Such pattern motion is not true
material rotation and therefore cannot be directly substituted for $\Omega_i$.
SASI may nevertheless influence spin-kick alignment indirectly, either by
shaping the ejecta mass asymmetry that later drives the tug-boat acceleration \citep{2017ApJ...837...84J} or
by producing a pre-explosion gravitational drag if the mode persists coherently for a sufficiently long time.


These considerations suggest that if the natal kick is produced mainly by the long-term gravitational tug-boat effect after shock revival, then rotation alone has a limited ability to enhance spin-kick alignment through the $\mathrm{sinc}(\beta/2)$ factor.  
However, if the NS were strongly magnetized and rapidly rotating, angular momentum could be transported outward to the Alfv\'en radius.  Increasing $\beta$ from $5\times10^{-3}$ to $5$ in this way requires $r_A\gtrsim30\,r_{NS}$ for sufficient angular momentum per unit mass of the NS  to be transported. The value of $r_A$ is likely to increase with time during the explosion, and if 
this catches up to locus of the tugging mass, then $\beta$ could increase with time. 

Even if $\beta$ is too low to affect the spin-kick alignment, rotation may alternatively affect spin-kick alignment  by somehow asymmetrizing the ejected mass distribution, represented by the $\tan{\theta}$ factor in Equation~(\ref{eq:main}). This possibility was also mentioned by \citet{2006A&A...457..963S}. Alternatively, magnetorotational jets in the explosions may  produce neutron-star kicks directly through asymmetric matter ejection,  
although  the situation remains unclear. For example, the 2D magnetorotational simulations of \cite{2024PhRvD.110h3025K} found kick velocities up to $\sim 500\,\mathrm{km\,s^{-1}}$, whereas the 3D magnetorotational simulations of \cite{2023MNRAS.522.6070P} found more modest kicks of about $150\,\mathrm{km\,s^{-1}}$ and no clear spin-kick alignment. There are also other proposed mechanisms \citep{2025RAA....25d5008B,2025ApJ...992..190S} involving stochastic, asymmetric jets without requiring rapid rotation.

\section{Conclusions}

In this paper, we have developed a  rotating extension of the gravitational tug-boat mechanism for neutron-star natal kicks. The model represents a 
mass asymmetry of the ejecta by an effective mass perturbation that expands outward while rotating around the spin axis. The main import of the model is that it identifies two possible ways in which rotation can affect spin-kick alignment from a kick derived from gravitational momentum transfer.  First, rotation can directly average away the transverse gravitational force; this effect is controlled by $\beta$.
Second, rotation can indirectly change the geometry of the explosion asymmetry;
this effect is contained in the distribution of $\theta$. Equation~(\ref{eq:main})
therefore provides a simple theoretical mapping between the rotating ejecta
asymmetry and the observable spin-kick angle.

The quantitative estimates of section~\ref{sec:implication} show that the sinc-factor effect is likely too weak to be influential for ordinary progenitor
rotation 
unless the NS is sufficiently magnetized and can transport additional angular momentum to the expanding shock region. 
As such, simple rotational averaging of the gravitational tug-boat force without  the influence of
 magnetic fields will not lead to a  strong spin-kick alignment in ordinary core-collapse supernovae. This may not necessarily disagree  with observations.
Current  evidence suggests some tendency toward spin-kick
alignment, but the inferred alignment is not that sharp. For example,
\cite{2012MNRAS.423.2736N} found evidence for projected spin-kick
alignment but argued that
the
underlying kick directions still require a broad angular dispersion,
$\sigma_v\sim30^\circ$. Thus, a successful theoretical model should not drive all kicks into nearly perfect alignment with the spin axis. A mechanism
that modestly biases the kick distribution toward smaller spin-kick angles, or
that modestly biases the underlying distribution of ejecta asymmetry relative to the
spin axis, may be sufficient.

Even if the sinc factor
in Equation~(\ref{eq:main}) is close to unity, rotation may still influence spin-kick alignment by changing the distribution of the effective polar angle $\theta$. For example, rotation
may modify neutrino-driven convection, affect the behavior of  instabilities such as SASI \citep{2017ApJ...837...84J}, or, in rapidly rotating magnetorotational explosions, produce asymmetric jet-like outflows (which may either change the mass asymmetry or act as a distinct kick mechanism).


Limitations of the model should be emphasized. We have represented the
ejecta asymmetry by a single effective mass perturbation with a fixed polar
angle and a prescribed radial expansion law. In reality, the gravitational force
on the proto-neutron star is produced by a turbulent, time-dependent,
three-dimensional mass distribution. The amplitude, direction, and coherence of
the dominant asymmetry can evolve during the explosion. 
Future  numerical simulations of rotating fluids may help to clarify how the asymmetric mass distribution evolves and gravitationally influences the neutron star.  We have also neglected details of 
neutrino transport,  feedback from
the neutron-star motion, detailed treatment of magnetic fields, the possible influence of binary interaction, and the nonlinear interaction between convection,
rotation, and ejecta expansion. 
All of these may be at work. Nevertheless, 
the simple model we have derived herein can help
in analyzing present and future state-of-the-art core-collapse simulations 
when determining the
amplitude, orientation, and angular motion of the dominant ejecta asymmetry
during the gravitational tug-boat phase,
 connecting the results 
to the observed
spin-kick alignment distribution of neutron stars.

We focused on the tug-boat kick not out of advocacy for that as necessarily being the dominant kick mechanism, but to study the influence of rotation on a well defined kick force. The calculations  can  be generalized for cases in which  the initial kick force comes from a different source. Our results suggest that for any kick force not sourced by the rotation itself, there will be some tendency toward spin-kick alignment and reduction of the kick magnitude, but the  quantitative strength of these effects will depend on the particular kick mechanism.

\section*{Acknowledgments}
We acknowledge support from the National Science Foundation (NSF), award no. PHY-2020249 which funds the NSF Physics  Center for Matter at Atomic Pressure (CMAP).  We thank N. Soker and A. Burrows for comments and discussions.

\bibliographystyle{plainnat}
\bibliography{references}

\end{document}